\documentclass[proceedings]{rmaa}
\usepackage{rmaacite}

\renewcommand{\P}[1]{%
\ifnum#1=1\hbox{OW~168--326E}\fi
\ifnum#1=2\hbox{OW~167--317}\fi
\ifnum#1=3\hbox{OW~163--317}\fi
\ifnum#1=5\hbox{OW~158--323}\fi
\ifnum#1=0\hbox{OW~171--334}\fi}

\title{Three-Dimensional Simulations of the Parker Instability in
a Uniformly-rotating Disk}
\author{Jongsoo Kim\altaffilmark{1},
        Dongsu Ryu\altaffilmark{2}, and 
        T. W. Jones\altaffilmark{3}}
\altaffiltext{1}{Korea Astronomy Observatory, 61-1, Hwaam-Dong, Yusong-Ku,
Taejon 305-348, Korea; jskim@hanul.issa.re.kr}
\altaffiltext{2}{Department of Astronomy and Space Science, Chungnam National 
University, Daejeon 305-764, Korea;\\ ryu@canopus.chungnam.ac.kr}
\altaffiltext{3}{Department of Astronomy, University of Minnesota, 
Minneapolis, MN 55455, USA; twj@astro.spa.umn.edu}

\shortauthor{Kim, Ryu \& Jones}
\shorttitle{Parker Instability in a Uniformly-rotating Disk}

\keywords{instabilities --- ISM: clouds --- ISM:
magnetic fields --- ISM: structure --- MHD}

\abstract{  

We investigate the nonlinear effects of uniform rotation on the Parker
instability in an exponentially-stratified disk through high-resolution
simulations.  During the linear stage, the speed of gas motion is subsonic
and the evolution with the rotation is not much different from that 
without the rotation. This is because the Coriolis force is small.
During the nonlinear stage, oppositely-directed supersonic flows near a
magnetic valley are under the influence of the Coriolis force with 
different directions, resulting in twisted magnetic field lines near the
valley.  Sheet-like structures, which are tilted with respect to the initial 
field direction, are formed with an 1.5 enhancement of column density with 
respect to its initial value.  Even though uniform rotation doesn't
give  much impact on density enhancement, it generates helically twisted 
field lines, which may become an additional support mechanism of clouds. 
}

\nonstopmode

\begin{document}

%% This command is necessary to typeset the title, abstract, etc. 
\maketitle

%%
%% And here starts the text....
%%
\section{Introduction}

It has been considered that the Parker instability (Parker 1966) is one 
of plausible mechanisms for the formation of the giant molecular
clouds (GMCs).  The conjecture is mainly based on the results from
linear analyses that the minimum growth time and its wavelength along 
the magnetic field direction are comparable to the life time and the
inter-distance of GMCs.  However, it is already known that the instability 
initiated by three-dimensional perturbations grows at the maximum growth
rate with infinite wavenumber along the third dimension perpendicular
to both the gravity and field directions (Parker 1967).  This means
that one dimension of structures formed by the instability should have a
very small scale, which is an obstacle for the GMC formation scenario. 

Linear analyses have inherent limitation that they are no longer valid in 
the nonlinear region.  In order to see the nonlinear effects, Kim et al.
(1998) followed up the evolution of the Parker instability without any
rotation through three-dimensional numerical simulations.  Sheet-like
structures formed at the developing stage of the instability persist
during the nonlinear stage, which confirms the above linear analyses result.
Furthermore, the density enhancement factor near the midplane 
was about 2, which is too small, and becomes another obstacle for the 
GMC formation scenario.     

It is also known through linear analyses (Shu 1974; Zweibel \& Kulsrud 1975)
that uniform rotation reduces the growth rate of the instability.  
It is the Colioils force that prevents the lateral motion and reduce
the growth rate. The purpose of this paper is to investigate the
nonlinear effects of uniform rotation on the Parker instability.  Based
on numerical simulations, we discuss structural deformation due to the
rotation and re-address the possibility of the GMC formation by the
Parker instability.

\section{Three-dimensional Simulations}

In order to describe the Parker instability at the solar neighborhood,
we introduce a Cartesian coordinate system $(x, y, z)$, whose axes are 
parallel to radial, azimuthal, and vertical directions of the Galaxy, 
respectively.  As a model for the initial distributions of gas and magnetic 
field, we choose a magnetohydrostatic equilibrium which was originally 
used by Parker (1966).  With downward gravity ($-g \hat{z}$) and 
horizontal magnetic field ($B_0[z] \hat{y}$), gas and magnetic pressures 
are described by an exponential function.  Its scale height is defined by 
$H=(1+\alpha)a^2/g$, where $\alpha$ is the ratio of magnetic to gas pressure 
and $a$ is the isothermal sound speed.  The solar neighborhood, as a whole,
is assumed to rotate with a constant angular speed ($\Omega \hat{z}$)
around the Galactic center. A computational cube, $0 \le x,y,z \le 12H$,
is placed at the solar neighborhood. To initiate the instability, random
velocity perturbations are added.  The standard deviation of each
velocity component is set to be $10^{-4}a$.  
A periodic condition at radial and azimuthal boundaries and 
a reflection condition at vertical boundaries are enforced.   
Isothermal MHD equations, which takes into account the
Coriolis force, are solved by the MHD TVD code (Kim et al. 1999).    
We have done high-resolution (256$^3$) simulations with
$\alpha=1$ and $\Omega a/(2H)=1$. In what follows, we use normalized
quantities with respect to $H$ and $a$. 

\begin{figure}[t]
  \begin{center}
    \leavevmode
    \includegraphics[width=0.49\textwidth]{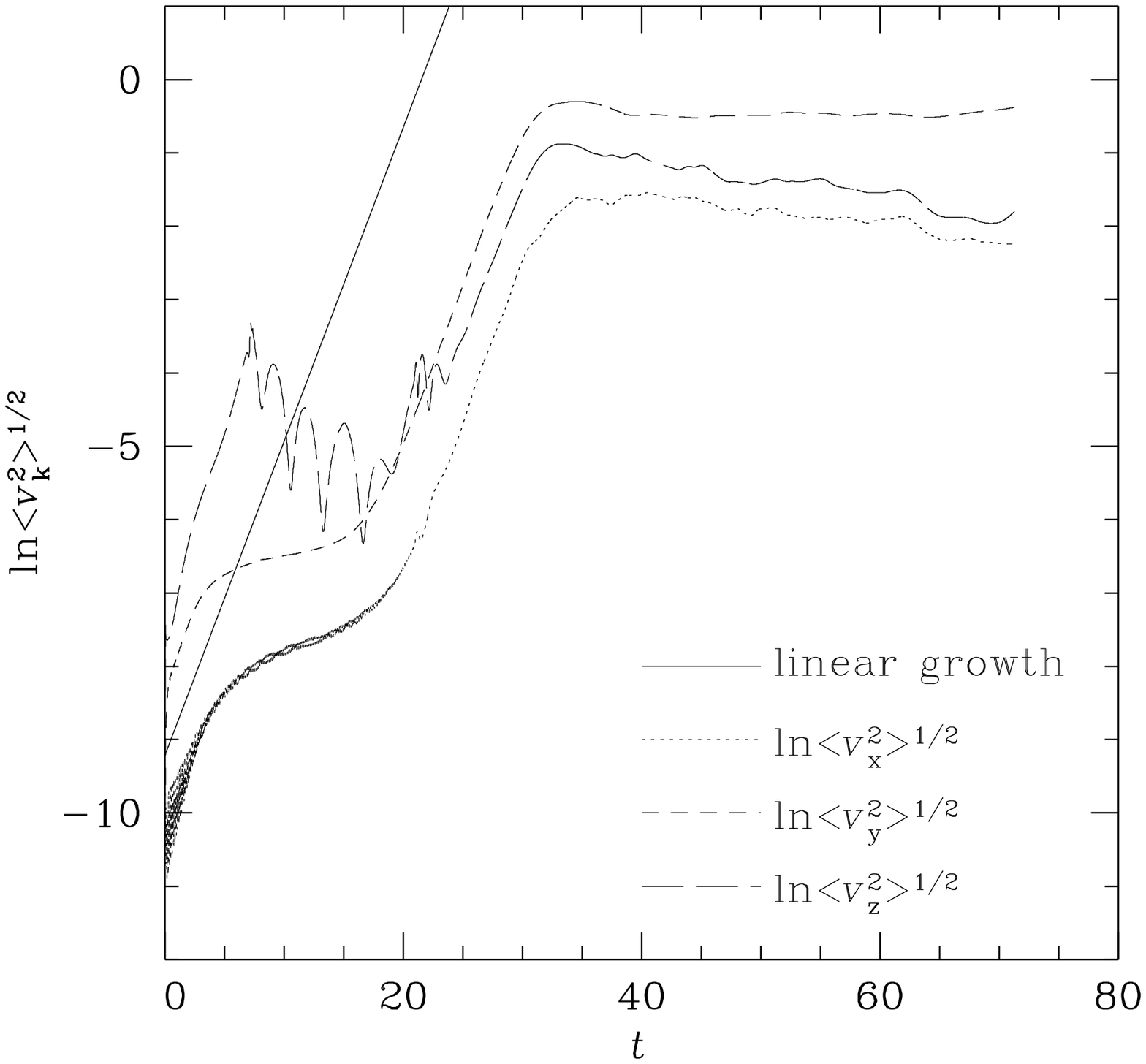} \hfil
    \includegraphics[width=0.49\textwidth]{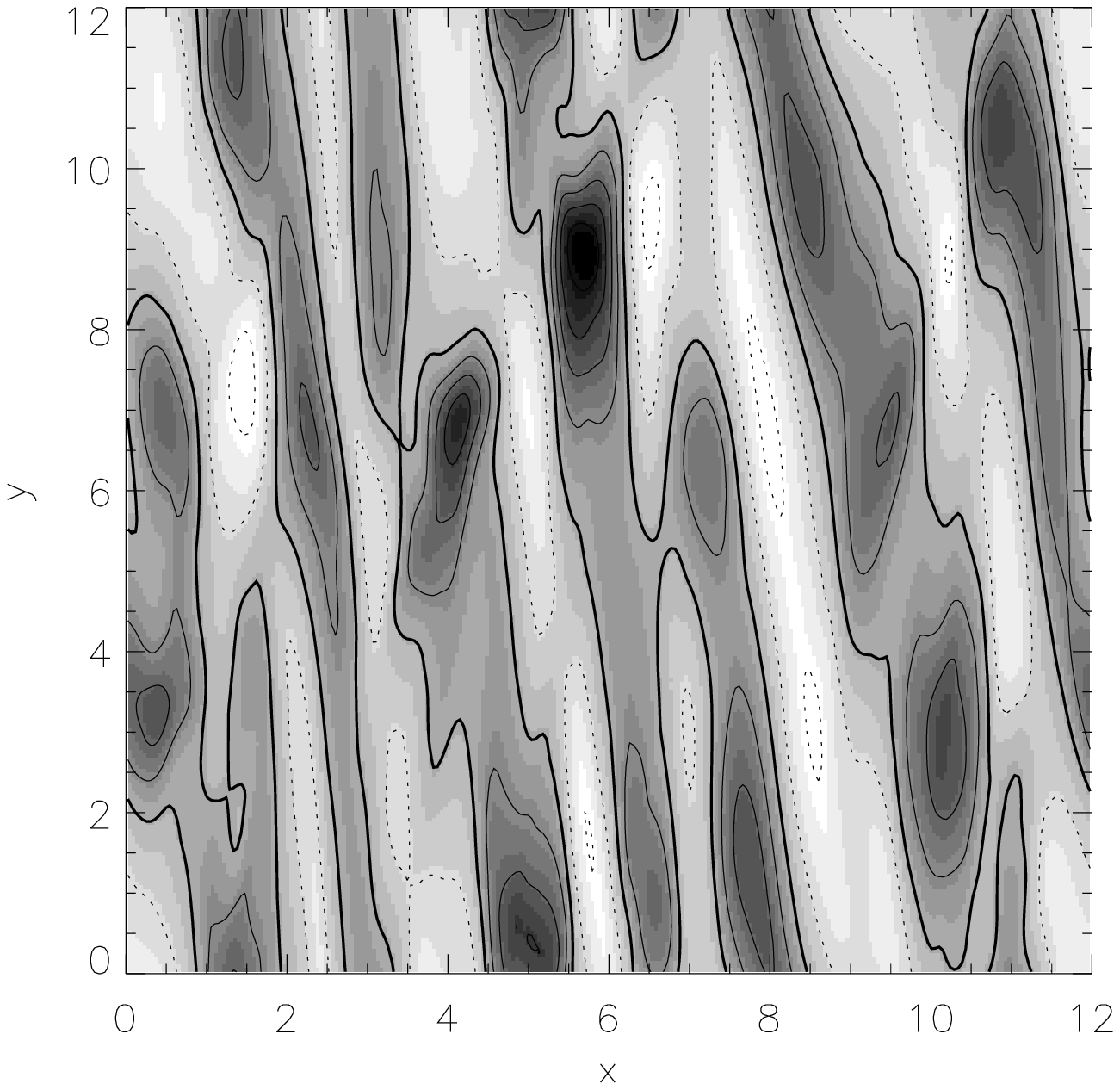}
\caption{(a) Rms of each velocity component as a function of time 
(left panel).  The maximum linear growth rate (0.41) of the Parker instability
in uniformly-rotating, exponentially-stratified disk is represented by a 
solid line.  (b) Normalized column density at $t=40$ (right panel; see 
eq.~[1]).  Gray-scales (where black represents the highest) and 
contours (0.8, 1.0, 1.2, 1.4) of column density are overlaid.         
The units of length, time and velocity are $H$, $H/a$ and 
$a$, respectively. 
}
  \end{center}
\end{figure}

If two-dimensional ($y$ and $z$) perturbations are considered, uniform
rotation reduces the growth rates of the undular mode (Zweibel \& Kulsrud
1975).  If the additional radial perturbations are given, not only 
the undular mode but also the interchange mode come into existence.  
When they work together, the mixed mode (undular + interchange)
is more unstable than the undular mode alone.  However, the interchange mode 
generates vertical motions, which are not affected by the rotation. 
For the extreme case with infinite radial wavenumber, the growth rates 
do not depend on angular speed.  The most unstable growth rate 
is 0.41 (Shu 1974).  Figure~1a represents the rms of each velocity
component as a function of time.  After the magnetized gas system adjusts
itself against the random perturbations, the rms velocities grow with
the maximum linear growth rate 0.41 from $t\simeq 20$ to $t\simeq 30$.  
When the rms of $y$-velocity is comparable to the isothermal sound 
speed ($t=30$), the evolution enters into the nonlinear stage.  

Normalized column density is defined by 
\begin{equation}
N(x,y;t) = \int_{0}^{12H} \rho(x,y,z;t) dz {\bigg /} \int_{0}^{12H} 
\rho_0(z) dz, 
\end{equation}
where $\rho_0(z)$ and $\rho(x,y,z;t)$ are the densities at initial and
specific time epochs. The column density at $t=40$ is shown
in Figure~1b. Sheet-like structures are formed, whose
long dimension is a bit tilted with respect to the initial field direction 
due to rotation.  The maximum enhancement of column density attains around 
$t\simeq 40$ and its value is about 1.5.  

Evolution of the Parker instability without rotation is described 
in Kim et al. (1998).  Here we mention briefly the effects of rotation.
Since the speed of gas motion during the linear stage is less than
the isothermal speed, the Coriolis force is not important and
the evolution with rotation is not much different from that without rotation.
As the instability fully develops during the nonlinear stage, 
magnetic arches and valleys form. The speed of falling gas toward magnetic 
valleys becomes comparable or larger than an isothermal speed.  Now the 
Coriolis force becomes important.  Oppositely-directed flows near the magnetic 
valleys are under the influence of the Coriolis force with different 
directions. Due to this reason, the magnetic field lines in the valley 
regions become twisted.  

In our previous study of the Parker instability without rotation 
(Kim et al. 1998), we concluded that it is difficult 
to regard the Parker instability alone as the formation mechanism of GMCs. 
The inclusion of rotation doesn't change this conclusion.  The density
structure is still sheet-like and the density enhancement factor is small.
Nevertheless, rotation generates helically twisted field lines,
which may become an additional support mechanism of clouds.

\acknowledgements
The work by JK was supported in part by the Office of the Prime Minister
through Korea Astronomy Observatory grant 99-1-200-00.
The work by DR was supported in part by KOSEF through grant and 981-0203-011-2.
The work by TWJ was supported in part by NSF grants AST9616964 and INT9511654, 
NASA grant NAGS-5055 and by the Minnesota Supercomputing Institute.

%% When using the rmaacite package, the \bibitem command should be of
%% the format: 
%%
%%             \bibitem[AUTHOR<YEAR>]{KEY} 
%%
%% so that the \cite{KEY}, etc. commands will work properly. 
%% 
%% If you are doing the citations manually, then you can just use
%% `\bibitem{}' instead. This will give you a warning about
%% `multiply-defined labels' which you can safely ignore.
%% 

\end{document}